\documentclass[12pt]{article}%
\usepackage{amsmath}
\usepackage{graphicx}
\usepackage{amsfonts}
\usepackage{amssymb}%
\providecommand{\U}[1]{\protect\rule{.1in}{.1in}}
\providecommand{\U}[1]{\protect\rule{.1in}{.1in}}
\providecommand{\U}[1]{\protect\rule{.1in}{.1in}}
\begin{document}

\title{Quantitative Calculation of the Spatial Extension of the Kondo Cloud}
\author{Gerd Bergmann\\Department of Physics\\University of Southern California\\Los Angeles, California 90089-0484\\e-mail: bergmann@usc.edu}
\date{\today}
\maketitle

\begin{abstract}
The internal s-electron polarization in the ground state of a Friedel-Anderson
and a Kondo impurity is calculated. The Wilson basis with exponentially fine
energies close to the Fermi level is used and expressed in terms of free
electron waves with linear energy-momentum dispersion. It is found that in the
singlet state the magnetic moment of the d-electron is screened by an
s-electron cloud. The linear extension of the cloud is inversely proportional
to the Kondo energy. When the singlet state is destroyed the polarization
cloud is absent. \newpage\ 

\end{abstract}

\section{Introduction}

The properties of magnetic impurities in a metal is one of the most
intensively studied problems in solid state physics. The work of Friedel
\cite{F28} and Anderson \cite{A31} laid the foundation to understand why some
transition-metal impurities form a local magnetic moment while others don't.
Kondo \cite{K8} showed that multiple scattering of conduction electrons by a
magnetic impurity yields a divergent contribution to the resistance in
perturbation theory. Kondo's paper stimulated a large body of theoretical and
experimental work which changed our understanding of d- and f-impurities
completely (see for example \cite{Y2}, \cite{V7}, \cite{S77}, \cite{D44},
\cite{H23}, \cite{M20}, \cite{A36}, \cite{G24}, \cite{C8}, \cite{H20}). A
large number of sophisticated methods were applied in the following three
decades to better understand and solve the Kondo and Friedel-Anderson
problems. In particular, it was shown that at zero temperature the
Friedel-Anderson impurity is in a non-magnetic state. To name a few of these
methods: scaling \cite{A51}, renormalization \cite{W18}, \cite{F30},
\cite{K58}, \cite{K59} Fermi-liquid theory \cite{N14}, \cite{N5}, slave-bosons
(see for example \cite{N7}), large-spin limit \cite{G19}, \cite{B103}. After
decades of research exact solutions of the Kondo and Friedel-Anderson
impurities were derived with help of the Bethe-ansatz \cite{W12}, \cite{A50},
\cite{S29}, representing a magnificent theoretical achievement. The
experimental and theoretical progress has been collected in a large number of
review articles \cite{D44}, \cite{H23}, \cite{M20}, \cite{A36}, \cite{G24},
\cite{C8}, \cite{H20}, \cite{W18}, \cite{N5}, \cite{N7}, \cite{B103},
\cite{W12}, \cite{A50}, \cite{S29}, \cite{N17}.

One of the most controversial aspects of the Kondo ground state is the
so-called Kondo cloud within the radius $\xi_{K}$ where $\xi_{K}$ is called
the Kondo length
\begin{equation}
\xi_{K}=\frac{\hbar v_{F}}{k_{B}T_{K}} \label{K_xi}%
\end{equation}
($k_{B}T_{K}$ = Kondo energy, $v_{F}$ = Fermi velocity of the s-electrons).

The idea is to divide the ground state $\Psi_{K}$ of a Kondo impurity into two
parts with opposite d-spins. (By reversing all spins one can transform one
component into the other one.) The proponents of the Kondo cloud argue that in
each component there is an s-electron within the Kondo sphere which
compensates the d-spin. This s-electron forms a singlet state with the d-spin.
An important assumption of the Kondo-cloud proponents is that, above the Kondo
temperature, the bond is broken and this screening cloud evaporates from the
Kondo sphere.

Already in the 1970's there were a number of theoretical papers with different
predictions about the Kondo cloud which stimulated several experimental
investigations. In most of the experiments nuclear magnetic resonance (NMR)
was used, for example in Cu samples with dilute Fe-Kondo impurities. (For
details see ref. \cite{S84} which contains also an overview of the theoretical
predictions at that time). By applying a magnetic field (small enough so that
it does not destroy the Kondo singlet state) NMR was used to measure the
electron spin polarization at shells of Cu atoms around the Fe impurity. The
line shift (adjusted with the temperature dependent susceptibility of the Fe
impurity) did not show any change when the temperature crossed the Kondo
temperature. This contradicted the concept that, in the Kondo ground state,
the d-impurity was paired with and screened by an s-electron which evaporates
above the Kondo temperature.

In recent theoretical papers the argument is made that an NMR experiment
cannot possibly detect the screening electron because of the large volume of
the Kondo sphere. With a radius of the order of $0.1\mu m$ (or larger) the
Kondo sphere contains more than $2.0\times10^{8}$ atoms. Then the Kondo cloud
is only a faint Kondo fog. It is spread so thinly that the change of
polarization felt by an individual Cu atom can not be detected in an NMR experiment.

Actually the experimental side has not improved so far but a number of
theoretical suggestions have been published since 2000 which propose to
observe the Kondo cloud in sub-micron structures, in particular in connection
with quantum dots \cite{A81}, \cite{A80}, \cite{A82}, \cite{B176}, \cite{S83}.
A number of these papers also contain numerical renormalization calculation to
obtain the correlation functions between the impurity spin and the spin of the
conduction electrons.

In this paper I calculate quantitatively the screening cloud of a
Friedel-Anderson and a Kondo impurity in the magnetic and the singlet state.

The paper is organized as follows. In section 2 the theoretical background of
this paper will be introduced, the FAIR method. In section 3 the numerical
results for the spatial density of the s-electrons will be reported and
discussed for several cases. First the spatial extent of the simple Friedel
d-resonance with spinless electrons is discussed. Since the physics of the
Friedel impurity is simple this consideration will give us confidence in the
use of the Wilson wave functions. In the next step the electronic environment
of the Friedel-Anderson (\textbf{FA}) impurity is investigated. The
calculation is first performed for its magnetic state. The latter represents
the magnetic building block of the singlet ground state of the FA-impurity
which is investigated next. Last, but not least, the ground state of the Kondo
Hamiltonian is analyzed. The subtle modification between the magnetic and the
singlet ground state is discussed. Section 4 contains the conclusion. In the
appendix details of the FAIR states and the Wilson states are presented,
including the wave functions of the Wilson states and their density in real space.%

\[
\]

\section{Theoretical Background}

\subsection{The FAIR Method}

To investigate the question of a Kondo cloud I first consider a
Friedel-Anderson impurity with spin 1/2 in a metal host. The underlying
Hamiltonian consists of two Friedel resonance Hamiltonians (one for each spin)
and a Coulomb term $Un_{d,\uparrow}n_{d,\downarrow}$.
\begin{equation}
H_{FA}=%
{\textstyle\sum_{\sigma}}
\left\{  \sum_{\nu=1}^{N}\varepsilon_{\nu}c_{\nu\sigma}^{\dag}c_{\nu\sigma
}+E_{d}d_{\sigma}^{\dag}d_{\sigma}+\sum_{\nu=1}^{N}V_{sd}(\nu)[d_{\sigma
}^{\dag}c_{\nu\sigma}+c_{\nu\sigma}^{\dag}d_{\sigma}]\right\}  +Un_{d,\uparrow
}n_{d,\downarrow} \label{hfa0}%
\end{equation}
Here $c_{\nu\sigma}^{\dag}$ represent the creation operators of the Wilson
s-electron state (which are discussed below), $d_{\sigma}^{\dag}$ represents
the creation operator of the d-state at the impurity and $n_{d,\sigma}$ are
the operators of the d-occupation.

In the appendix I briefly describe the development of a very compact solution
for the magnetic state and the singlet state. It uses two localized s-states
$a_{0+}^{\dag}$ and $a_{0-}^{\dag}$ as artificial Friedel resonance states or
\textbf{FAIR }states (Friedel Artificially Inserted Resonance states). The
ground state solutions in terms of the FAIR basis have been studied in recent
years for the Friedel Hamiltonian \cite{B91}, \cite{B92}, the magnetic and the
singlet state of the Friedel-Anderson Hamiltonian \cite{B152}, \cite{B151} and
the Kondo Hamiltonian \cite{B153}. The compact and explicit form makes these
solutions ideal for calculation of spatial properties. The compact ground
states which are used in this paper are given for the Friedel impurity by equ.
(\ref{Psi_F}), the magnetic state by equ. (\ref{Psi_MS}), the singlet state of
the FA impurity by equ. (\ref{Psi_SS}) and the Kondo ground state by equ.
(\ref{Psi_K}).

The potentially magnetic state $\Psi_{MS}$ of the Friedel-Anderson Hamiltonian
(which is the building block for the singlet state $\Psi_{SS}$) has the
following form
\begin{equation}
\Psi_{MS}=\left[  A_{s,s}a_{0-\downarrow}^{\ast}a_{0+\uparrow}^{\ast}%
+A_{d,s}d_{\downarrow}^{\ast}a_{0+\uparrow}^{\ast}+A_{s,d}a_{0-\downarrow
}^{\ast}d_{\uparrow}^{\ast}+A_{d,d}d_{\downarrow}^{\ast}d_{\uparrow}^{\ast
}\right]  \prod_{i=1}^{n-1}a_{i-\downarrow}^{\ast}\prod_{i=1}^{n-1}%
a_{i+\uparrow}^{\ast}\Phi_{0}\label{Psi_MS}%
\end{equation}

The magnetic state $\Psi_{MS}$ has the same form as the mean field solution
$\Psi_{mf}$. The only difference is that the state $\Psi_{MS}$ opens a wide
playing field for optimization: (i) The FAIR states $a_{0+}^{\dag}$ and
$a_{0-}^{\dag}$ can be individually optimized, each one defining a whole basis
$\left\{  a_{i\pm}^{\dag}\right\}  $ (which yields a Hamiltonian of the form
in Fig.10a in the appendix) and (ii) the coefficients $A_{\alpha,\beta}$ can
be optimized as well, fulfilling only the normalization condition $A_{s,s}%
^{2}+A_{d,s}^{2}+A_{s,d}^{2}+A_{d,d}^{2}=1$. This yields a good treatment of
the correlation effects. The optimization procedure is described in detail in
ref. \cite{B152}. The state is denoted as the (potentially) magnetic state
$\Psi_{MS}$. It is expected to be a good solution above the Kondo temperature.

The Coulomb interaction in the Friedel-Anderson impurity destroys the symmetry
between the occupation of spin up and down electrons and generates a magnetic
moment (if $U$ is large enough). However, this state with broken symmetry is
degenerate. Its counterpart with reversed spins has the same energy, and at
zero temperature the two form a new symmetric state which has a different
symmetry than original one. This is somewhat analogous to an atom in a
harmonic potential. If some process transforms the harmonic potential into a
double well potential the symmetry is broken. Although the atom has now an
energy minimum on the left or right side its atomic wave function creates a
new symmetric state by forming a symmetric superposition of left and right occupation.

In a similar way I use the magnetic state as a building block for the singlet
ground state. $\Psi_{MS}$ together with its counterpart\ where all spins are
reversed yield two states $\overline{\Psi}_{MS}\left(  \uparrow\downarrow
\right)  $ and $\overline{\overline{\Psi}}_{MS}\left(  \downarrow
\uparrow\right)  .$ The singlet ground state is then given by
\[
\Psi_{SS}=\overline{\Psi}_{MS}\left(  \uparrow\downarrow\right)  \pm
\overline{\overline{\Psi}}_{MS}\left(  \downarrow\uparrow\right)
\]%
\begin{align}
\Psi_{SS} &  =\left[  \overline{A_{s,s}}a_{0-\downarrow}^{\ast}a_{0+\uparrow
}^{\ast}+\overline{A_{d,s}}d_{\downarrow}^{\ast}a_{0+\uparrow}^{\ast
}+\overline{A_{s,d}}a_{0-\downarrow}^{\ast}d_{\uparrow}^{\ast}+\overline
{A_{d,d}}d_{\downarrow}^{\ast}d_{\uparrow}^{\ast}\right]  \prod_{i=1}%
^{n-1}a_{i-\downarrow}^{\ast}\prod_{i=1}^{n-1}a_{i+\uparrow}^{\ast}\Phi
_{0}\label{Psi_SS}\\
&  +\left[  \overline{\overline{A_{s,s}}}a_{0+\downarrow}^{\ast}a_{0-\uparrow
}^{\ast}+\overline{\overline{A_{d,s}}}a_{0+\downarrow}^{\ast}d_{\uparrow
}^{\ast}+\overline{\overline{A_{s,d}}}d_{\downarrow}^{\ast}a_{0-\uparrow
}^{\ast}+\overline{\overline{A_{d,d}}}d_{\downarrow}^{\ast}d_{\uparrow}^{\ast
}\right]  \prod_{i=1}^{n-1}a_{i+\downarrow}^{\ast}\prod_{i=1}^{n-1}%
a_{i-\uparrow}^{\ast}\Phi_{0}\nonumber
\end{align}
(If all spin up states are moved to the left the plus sign applies). The bars
on top of $\Psi_{MS}$ are supposed to point out that all the parameters such
as $a_{0\pm}^{\dag}$ and $A_{\alpha,\beta}$ have to be newly optimized.

The resulting ground state is a very good approximation for the exact ground
state. Its ground-state energy and the d-occupations for zero, one and two
d-electrons at the impurity are of the same quality as the best numerical
calculations in the field \cite{G34}.

In the numerical construction of the FAIR ground state the states $a_{0\sigma
}^{\dag}$ is rotated in the $N$-dimensional Hilbert space until the energy
expectation value reaches its minimum. Each time one has to construct the
other states of the basis $\left\{  a_{i\sigma}^{\dag}\right\}  $ so that they
are orthonormal and non-interacting. This cannot be done with $10^{23}$
electron states. Therefore we follow Wilson in constructing a reduced basis,
in our case with generally $N=50$ Wilson states. The definition of the Wilson
states, their properties and in particular their wave function in real space
is discussed in the appendix.

\section{Numerical Results}

\subsection{Friedel Impurity}

I start the analysis with the Friedel resonance of a spinless d-impurity.
Since the Friedel Hamiltonian is a single particle Hamiltonian which can be
solved exactly, one is here on familiar ground and the interpretation of the
results will guide us in the less known territory of the Friedel-Anderson and
the Kondo impurity.

Given is the $n$-particle free electron state $\Psi_{0}^{\left(  n\right)  }=%
{\textstyle\prod\limits_{\nu=0}^{n-1}}
c_{\nu}^{\dag}\Phi_{0}$ in terms of Wilson states. A (spinless) Friedel
d-impurity is introduced at the position $x=0$. This impurity has an
s-d-interaction with the states $\psi_{\nu}\left(  x\right)  =\left\langle
c_{\nu}^{\dag}\Phi_{0}|x\right\rangle .$ Due to the impurity the $n$-electron
state $\Psi_{0}^{\left(  n\right)  }$ is modified. As discussed in the
appendix the exact ground state of the Friedel impurity can be written as
\begin{equation}
\Psi_{F}=\left(  Aa_{0}^{\dag}+Bd^{\dag}\right)
{\textstyle\prod\limits_{i=1}^{n-1}}
a_{i}^{\dag}\Phi_{0}\label{Psi_F}%
\end{equation}
The composition of the FAIR state $a_{0}^{\dag}$ and the coefficients $A,B$
are obtained by an iteration process in which the FAIR state is rotated in
Hilbert space into its optimal orientation. Details of this iteration process
and the construction of the full ground state have been described in previous
papers \cite{B91}, \cite{B151}.

In the next step we have to calculate the integrated electron density in the
presence of the Friedel d-impurity at the origin. This calculation is
discussed in the appendix. The density of the state $\psi_{\nu}\left(
x\right)  $ is given by%
\[
\rho_{\nu}^{0}\left(  x\right)  =\left\vert \psi_{\nu}\left(  x\right)
\right\vert ^{2}=2^{\nu+3}\frac{\sin^{2}\left(  \pi x\frac{1}{2^{\nu+2}%
}\right)  }{\pi x}dx
\]
for $\nu<N/2.$ In the evaluation we use $N=50$ Wilson states. From the above
equation one recognizes that the density $\rho_{\nu}^{0}$ of the state
$\psi_{\nu}$ lies roughly in the interval between $2^{\nu-2}$ and $2^{\nu+2}$
(in units of $\lambda_{F}/2$). Since for negative energies $\nu$ takes the
values from $0\ $to $\left(  N/2-1\right)  $ the different wave functions
$\psi_{\nu}\left(  x\right)  $ have very different spatial ranges and
therefore very different densities, the lowest being of the order of
$2^{-25}<3\times10^{-8}$. Therefore one has to be careful in the summation of
the different contributions.

In the numerical evaluation we calculate (i) the s-electron density in the
presence and absence of the d-impurity, (ii) form the net s-electron density
$\rho\left(  x\right)  $ as the difference, (iii) integrate the net s-electron
density from $x=0$ to $r$ where $r$ is increased on an exponential scale,
$r=2^{l}$ and plot the net integrated s-electron density versus $l=\log
_{2}\left(  r\right)  $. This net integrated density $q\left(  r\right)
=\int_{0}^{r}\rho\left(  x\right)  dx$ is plotted in Fig.1 as a function of
the distance from the impurity. The full triangles and circles\ give the
integrated net s-electron density for the two components $\Psi_{A}=%
{\textstyle\prod\limits_{i=0}^{n-1}}
a_{i}^{\dag}\Phi_{0}$ and $\Psi_{B}=d^{\dag}%
{\textstyle\prod\limits_{i=1}^{n-1}}
a_{i}^{\dag}\Phi_{0}$ of the ground state, and the squares for the full ground
state $\Psi_{F}$. The curve for $\Psi_{A}$ increases from $0$ to $0.75$
roughly between $l=-2$ and $+3$. Around $l=23$ it drops back to zero. It has
to return to zero because the total number of occupied states in $\Psi_{A}$ is
the same as before in the state $\Psi_{0}=%
{\textstyle\prod\limits_{i=0}^{n-1}}
c_{\nu}^{\dag}\Phi_{0},$ i.e.\ equal to $N/2$. The equivalent distance
$r=2^{23}$ corresponds roughly to the maximum range of the wave functions. In
a way it can be considered as the border or surface of the theoretical sample.
Therefore the interpretation of this curve in Fig.1 is that in the state
$\Psi_{A}$ a fraction of $0.75$ electrons has moved from the surface towards
the d-impurity. There it surrounds the empty d-state in the spatial region
between $r=2^{-2}$ and $2^{3}$. (The component $\Psi_{A}$ has no d-electron).
The behavior of the integrated net density beyond about $l\thickapprox20$ is a
surface effect and of no interest for the present investigation. If one
increases the number of cells from $N$ to $N^{\prime}$ then this part of the
curve moves\ by $\left(  N^{\prime}-N\right)  /2$ to the right while the left
part of the curve remains unchanged.

The full circles in Fig.1 give the change of the net integrated density of
s-electrons for the state $\Psi_{B}$ as a function of distance. At large
distances ($r\thickapprox2^{23}$) it approaches the value $-1$ because this
state has only $\left(  N/2-1\right)  $ s-electrons. Between the distances
$l=-2$ and $+3$ the curve for $\Psi_{B}$ assumes the value $-.25$. If one adds
to the integrated s-electron density the one d-electron at the origin then one
obtains the dotted curve which starts on the left at one and approaches zero
at large distances. It is interesting to notice that in the range $5<l<20$ the
total integrated density of d- and s-electrons is equal to $0.75$ for both
wave functions $\Psi_{A}$ and $\Psi_{B}$. This means both components have
accumulated the same charge close to the impurity. As discussed below this
corresponds to a Friedel phase shift of $\delta=0.75\pi$.

In this specific example the parameters of $\left\vert V_{sd}^{0}\right\vert
^{2}=0.1$ and $E_{d}=-.135$ were chosen so that the occupation numbers take
the values $A^{2}=.25$ and $B^{2}=.75$. Then the net integrated density of
$\Psi_{A}$ times the occupation number $A^{2}$ yields $0.75\ast
0.25\thickapprox\allowbreak0.188$ and just cancels the corresponding product
of $0.75$*$\left(  -.25\right)  $ for the state $\Psi_{B}$. As a result the
total integrated density of s-electrons as a function of distance from the
impurity vanishes as the curve for $\Psi_{F}$ shows (full squares). There is
no change of s-electron density about the impurity. On the other hand the
curve approaches the value of $-.75$ for large distance. This is the charge
which has been moved from the surface into the d-state. This charge is
predicted by the Friedel sum rule. The occupation $B^{2}=0.75$ of the state
$\Psi_{B}$ means that the d-occupation is $0.75,$ i.e. the phase shift is
$\delta=.75\pi$ and therefore the Friedel charge is $\delta/\pi=0.75$.%

\begin{align*}
&
{\includegraphics[
height=3.3997in,
width=4.0324in
]%
{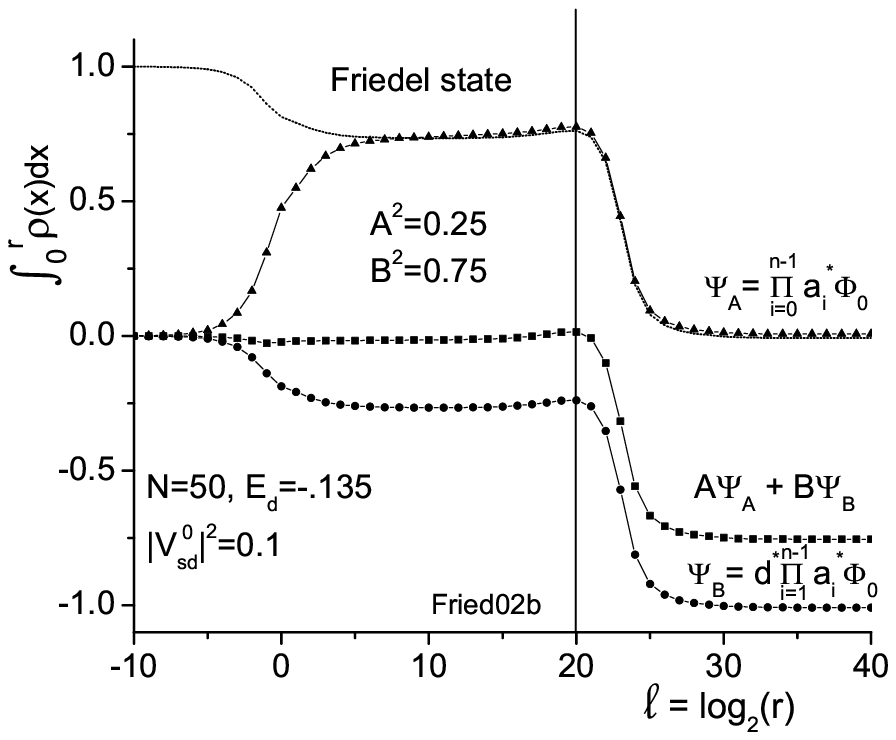}%
}%
\\
&
\begin{tabular}
[c]{l}%
Fig.1: The net integrated density $\int_{0}^{r}\rho\left(  x\right)  dx$ is
plotted versus the logarithm\\
of the distance $l=\log_{2}\left(  r\right)  $ from the impurity for the
components $\Psi_{A},\Psi_{B}$\\
and the full wave function $\Psi_{F}$. For the dotted line the additional
one\\
d-electron has been added.
\end{tabular}
\end{align*}

The Friedel resonance is also ideal testing ground for the spatial extension
of the wave function. For that purpose the d-state is moved to the Fermi level
($E_{d}=0)$. The half width of the resonance is given by $\Delta=\pi\left\vert
V_{sd}^{0}\right\vert ^{2}\rho_{0}$ ($\rho_{0}$=density of states). This
resonance width introduces an additional length scale into the problem. In the
resonance the d-state hybridizes with the s-electrons in this energy range of
2$\Delta.$This corresponds to a wave number range of $k_{\Delta}%
\thickapprox2\pi\left\vert V_{sd}^{0}\right\vert ^{2}\rho_{0}$ and therefore a
smearing of the wave function over a range $x_{\Delta}\varpropto1/\left(
2\pi\left\vert V_{sd}^{0}\right\vert ^{2}\rho_{0}\right)  $ ($V_{sd}^{0}$ is
measured in unit of $\varepsilon_{F}$ and $x_{\Delta}$ in units $\lambda
_{F}/2$).

In the numerical evaluation I change the value of $\left\vert V_{sd}%
^{0}\right\vert ^{2}$ in steps of two from $0.005$ to $0.16$. Indeed one
recognizes in Fig.2 that with decreasing $\left\vert V_{sd}^{0}\right\vert
^{2}$ the increase of the integrated\ net density is stretched over an
increasing length scale. Since the horizontal separation of the points by 1
corresponds to a factor of two in the length scale each curve is broader by a
factor two than the previous one.%
\begin{align*}
&
{\includegraphics[
height=2.4491in,
width=3.1374in
]%
{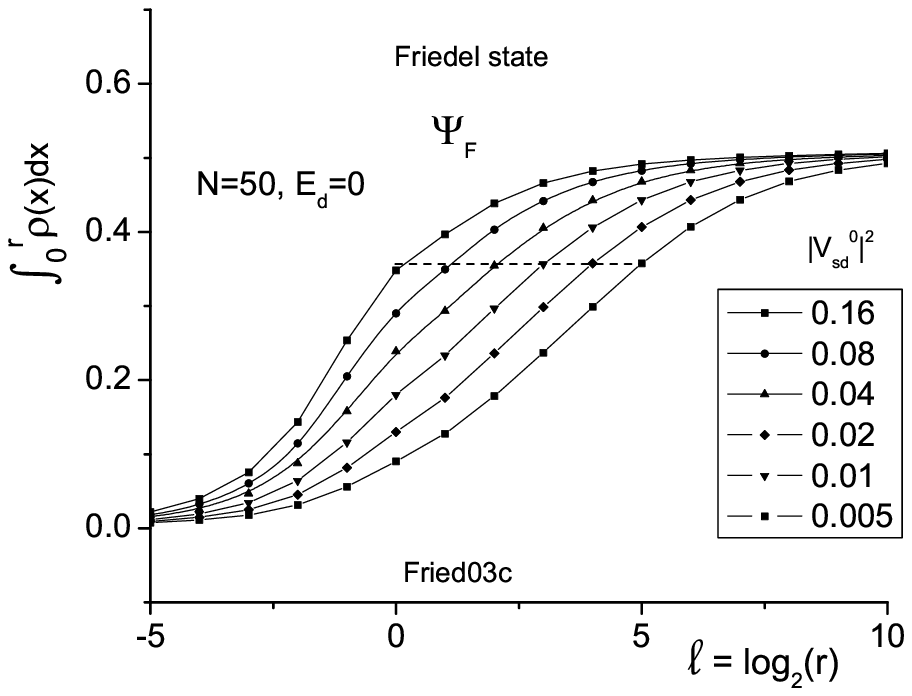}%
}%
\\
&
\begin{tabular}
[c]{l}%
Fig.2: The net integrated density $\int_{0}^{r}\rho\left(  x\right)  dx$ for
the state $\Psi_{A}$=$%
{\textstyle\prod\limits_{i=0}^{n-1}}
a_{i}^{\dag}\Phi_{0}$ for different values of\\
$\left\vert V_{sd}^{0}\right\vert ^{2}$ while the d-resonance lies at the
Fermi level, $E_{d}=0$. The width of the wave\\
function increases proportional to $1/\pi\left\vert V_{sd}^{0}\right\vert
^{2}$.
\end{tabular}
\end{align*}

\subsection{The Magnetic State}

The magnetic state $\Psi_{MS}$ is the building block of the singlet state. Its
multi-electron state is built from four Slater states and shown in equ.
(\ref{Psi_MS}). I calculate the net integrated density of spin up and down, as
well as total density and spin polarization in the vicinity of the impurity
when it is in the magnetic state. For the present purpose I choose the
parameters $E_{d}=0.5,$ $\left\vert V_{sd}^{0}\right\vert ^{2}=0.04,$ $U=1$
and $N=50$. This yields a well developed magnetic moment of $\mu=0.93\mu_{B}$.
The occupation of the different components is $A_{s,s}^{2}=0.0294,$
$A_{s,d}^{2}=0.0057,$ $A_{d,s}^{2}=0.9355$ and $A_{d,d}^{2}=0.0294.$

In Fig.3 the net densities are plotted. As discussed above, the range beyond
$2^{20}$ corresponds to the rim or surface of the sample and is of no interest
for the density distribution around the impurity. One recognizes that there is
only a negligible polarization of the electron gas in the vicinity of the
impurity. The important result of Fig.3 is that there is no polarization cloud
around the magnetic state of the impurity.
\begin{align*}
&
{\includegraphics[
height=3.2893in,
width=4.0872in
]%
{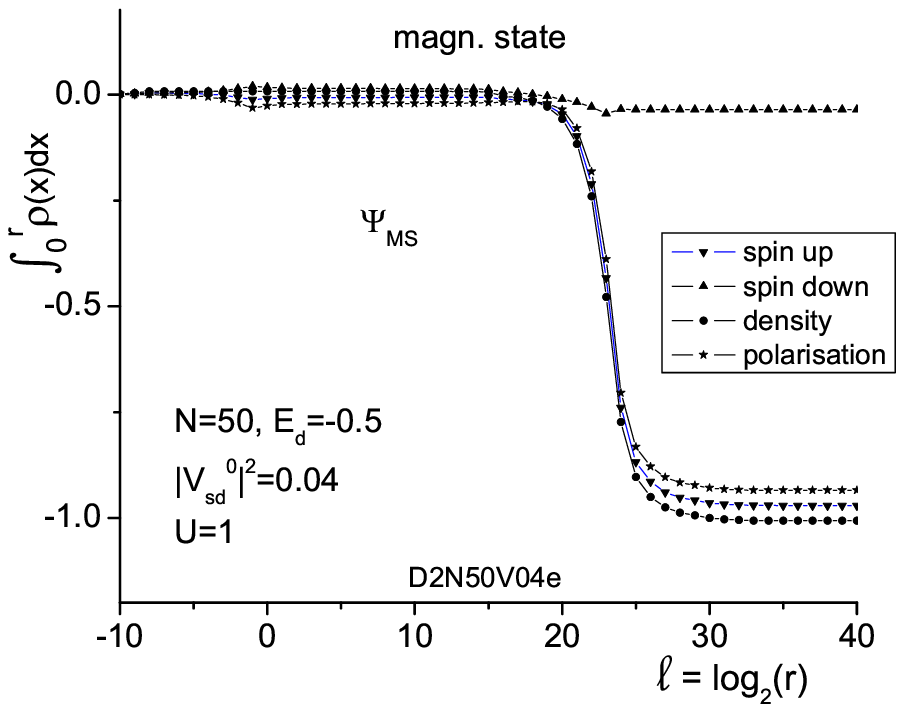}%
}%
\\
&
\begin{tabular}
[c]{l}%
Fig.3: The net integrated density $\int_{0}^{r}\rho\left(  x\right)  dx$ of
the s-electron within a distance $r$\\
from the impurity for spin up and down, as well as total density and spin\\
polarization. The magnetic moment of the impurity is $0.93\mu_{B}$.
\end{tabular}
\end{align*}

\subsection{The Singlet State of the Friedel-Anderson Impurity}

The author derived the compact singlet ground state in ref. \cite{B151}. Its
form is given in equ. (\ref{Psi_SS}). The singlet state does not have a net
spin polarization because of the symmetry between spin up and down. However,
the individual components may possess a spin polarization cloud. To
investigate this question one has first to decide how one subdivides the
symmetric state of the impurity. Since there are four Slater states which have
either zero or two d-electrons one cannot divide the full state into two parts
in which either the $d_{\uparrow}$- or the $d_{\downarrow}$-state are
occupied. A reasonable division would be into the states $\overline{\Psi
_{MS}\left(  \uparrow\downarrow\right)  }$ and $\overline{\overline{\Psi
_{MS}\left(  \downarrow\uparrow\right)  }}$. However, one may object that both
states contain a component in which, for example, the $d_{\uparrow}$-state is
singly occupied. Nevertheless I use the components $\overline{\Psi_{MS}\left(
\uparrow\downarrow\right)  }$ as the magnetic component because in the singlet
state the coefficients $\overline{A_{s,s}},..\overline{A_{d,d}}$ and
$\overline{\overline{A_{s,s}}},..\overline{\overline{A_{d,d}}}$ have the same
sign while in the triplet state they have opposite signs. (Their absolute
values are in both cases pair-wise identical).

In the following plots in Fig.4 I choose the state $\overline{\Psi_{MS}\left(
\uparrow\downarrow\right)  }$ with the majority d-spin pointing up. The same
parameters as for the magnetic state are used: $E_{d}=-0.5,$ $\left\vert
V_{sd}^{0}\right\vert ^{2}=0.04$, $U=1$. This yields the following
coefficients: $A_{ss}^{2}=0.0146,$ $A_{sd}^{2}=0.0028,$ $A_{ds}^{2}=0.4629$
and $A_{dd}^{2}=0.0146$. They are about half the values of the magnetic state.
Therefore on a superficial glance it appears that $\Psi_{ss}\thickapprox
\left(  1/\sqrt{2}\right)  \left[  \Psi_{MS}\left(  \uparrow\downarrow\right)
+\Psi_{MS}\left(  \downarrow\uparrow\right)  \right]  .$

In Fig.4 the integrated densities of spin up and down electrons, their sum and
difference (the polarization) are plotted versus the distance from the
magnetic impurity (on a logarithmic scale). One recognizes that now one has
considerable contributions to the integrated net densities of both spins. The
polarization of the two contributions is no longer zero but reaches a value of
$-0.46$ at a distance of $r\thickapprox2^{1\text{1.6}}$. Since the magnetic
state with net d-spin up has only a weight of about $1/2$ it contributes an
effective $d_{\uparrow}^{\dag}$-moment of $0.93/2\thickapprox0.46$. Therefore
this d-spin is well compensated by the polarization of the s-electron
background. The difference with the pure magnetic state is particularly
striking. We observe a screening polarization cloud of s-electrons about the
impurity within the range of $r\thickapprox2^{1\text{1.6}}=\allowbreak
3.1\times10^{3}$.%
\begin{align*}
&
{\includegraphics[
height=3.4147in,
width=4.1735in
]%
{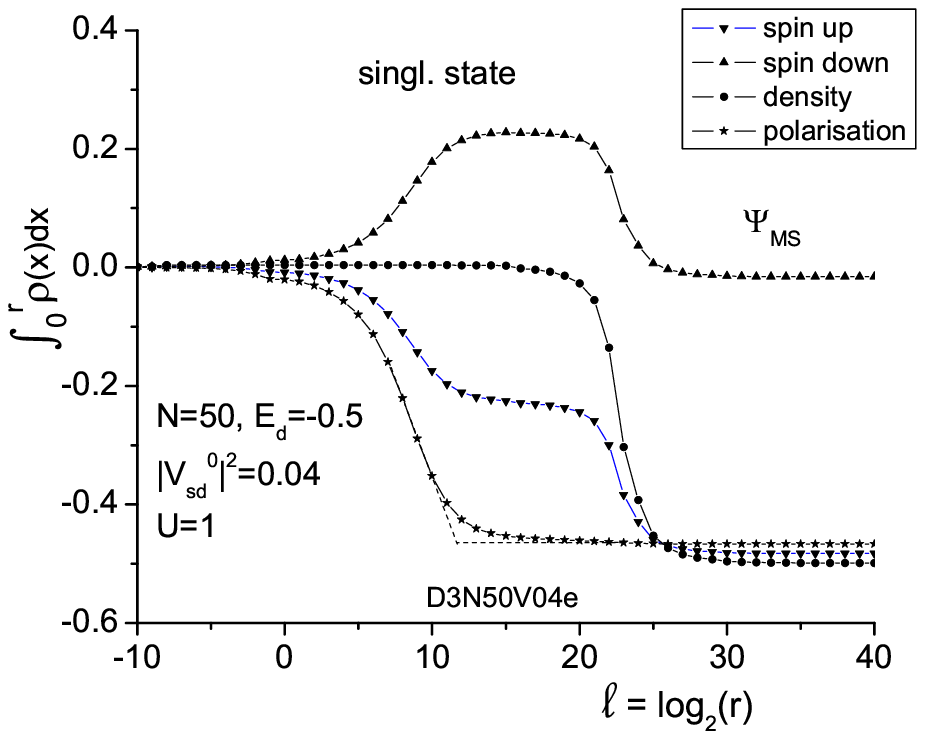}%
}%
\\
&
\begin{tabular}
[c]{l}%
Fig.4: The net integrated density $\int_{0}^{r}\rho\left(  x\right)  dx$
within a distance $r=2^{l}$\\
from the d-spin up component of the impurity$.$ Shown are the spin up, spin\\
down components as well as the total density and the polarization. The
d$_{\uparrow}$-spin\\
of $0.93/2$ is screened by 0.46 s-electrons within the range of $r\thickapprox
2^{1\text{1.6}}$.
\end{tabular}
\end{align*}

The range of the screening cloud depends on the strength of $\left\vert
V_{sd}^{0}\right\vert ^{2}$. In Fig.5 a similar plot as in Fig.4 is performed
for an s-d-coupling of $\ \left\vert V_{sd}^{0}\right\vert ^{2}=0.1$. This
time the polarization cloud extends only over a distance of $r\thickapprox
2^{6.4}\thickapprox\allowbreak84$.
\begin{align*}
&
{\includegraphics[
height=3.3997in,
width=4.2142in
]%
{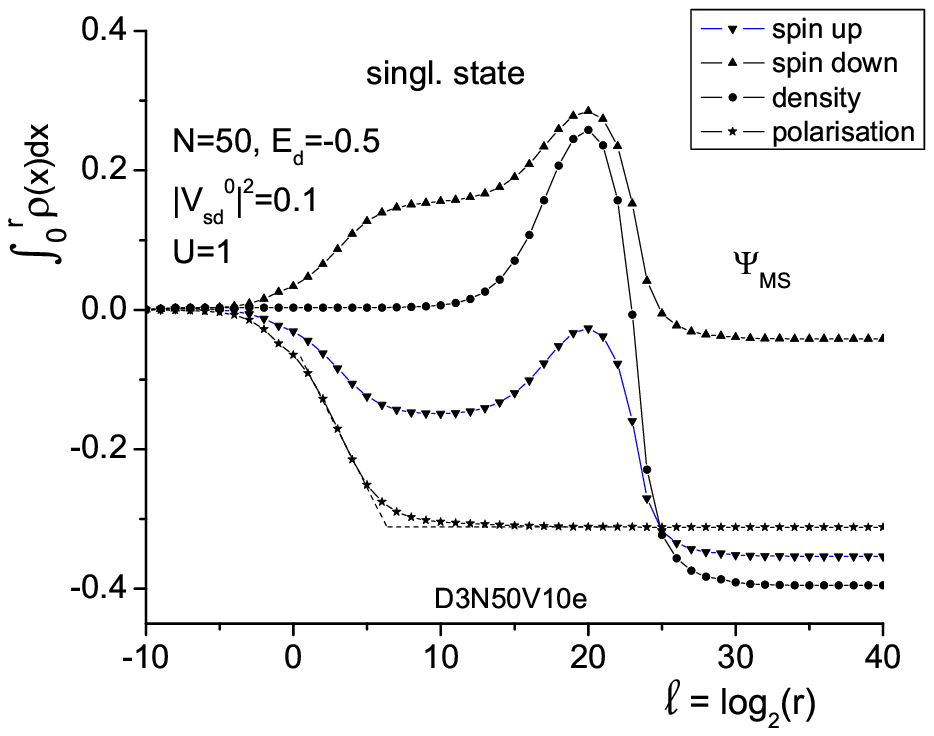}%
}%
\\
&
\begin{tabular}
[c]{l}%
Fig.5: The net integrated density $\int_{0}^{r}\rho\left(  x\right)  dx$
within a distance $r=2^{l}$\\
form the d-spin up component of the impurity$.$ Shown are the spin up, spin\\
down components as well as the total density and the polarization. The
d$_{\uparrow}$-spin\\
polarization cloud extends for $\left\vert V_{sd}^{0}\right\vert ^{2}=0.1$
over a range of $r\thickapprox2^{6.4}$.
\end{tabular}
\end{align*}

\subsection{The FAIR states}

It is remarkable how different the polarizations of the s-electrons in the
magnetic state $\Psi_{MS}$ and the magnetic component of the singlet state
$\overline{\Psi_{MS}}$ are. This is particularly surprising since the
amplitudes $\overline{A_{\alpha,\beta}}=\overline{\overline{A_{\alpha,\beta}}%
}$ in the singlet state are roughly equal to $A_{\alpha,\beta}/\sqrt{2}$.
However, in the singlet state one has a finite coupling between $\overline
{\Psi_{MS}}$ and $\overline{\overline{\Psi_{MS}}}$. This affects the
composition of the FAIR states $a_{0+}^{\dag}$ and $a_{0-}^{\dag}$. The two
states $a_{0+}^{\dag}$ and $a_{0-}^{\dag}$ contain the whole information about
the many electron states $\Psi_{MS}$ or $\Psi_{SS}$. When $a_{0\pm}^{\dag}$
are known the whole bases $\left\{  a_{i+}^{\dag}\right\}  $and $\left\{
a_{i-}^{\dag}\right\}  $ and the coefficients $A_{\alpha,\beta}$ can be
reconstructed. Therefore it is worthwhile to compare the two FAIR states for
the magnetic and the singlet state. 

For the magnetic state the coefficients $\alpha_{0\pm}^{\nu}$ of the states
$a_{0\pm}^{\dag}=%
{\textstyle\sum_{\nu}}
\alpha_{0\pm}c_{\nu}^{\dag}$ are plotted in Fig.6 versus the cell number
$\nu,$ which is a measure of the energy. While $a_{0+}^{\dag}$ is essentially
concentrated at positive energies the coefficients of $a_{0-}^{\dag}$ have
their main weight at negative energies. The two are mirror images with respect
to the energy zero.%

\begin{align*}
&
\begin{array}
[c]{cc}%
{\includegraphics[
height=2.1519in,
width=2.7347in
]%
{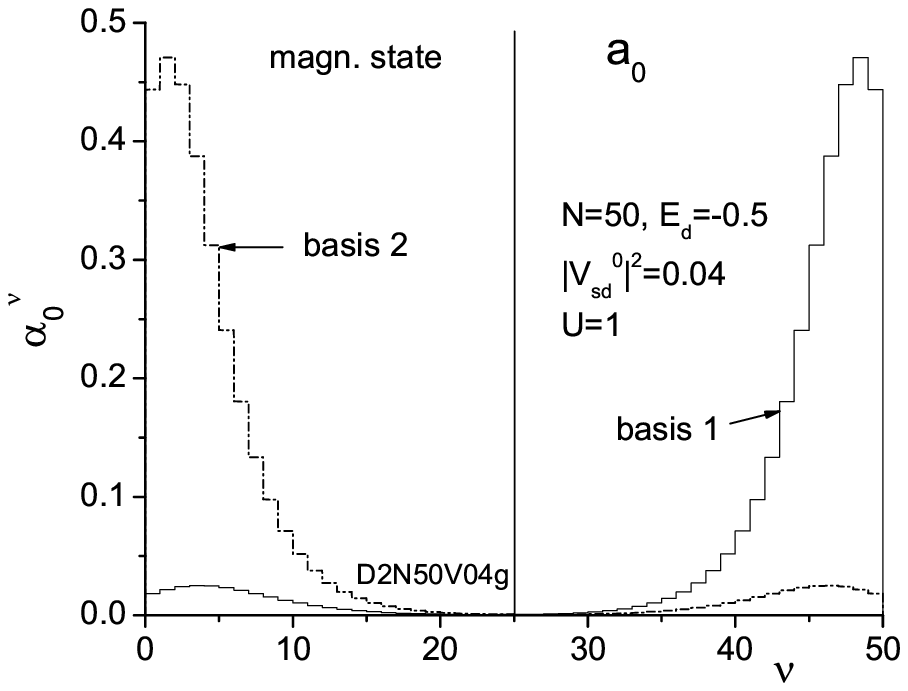}%
}%
&
{\includegraphics[
height=2.1162in,
width=2.6866in
]%
{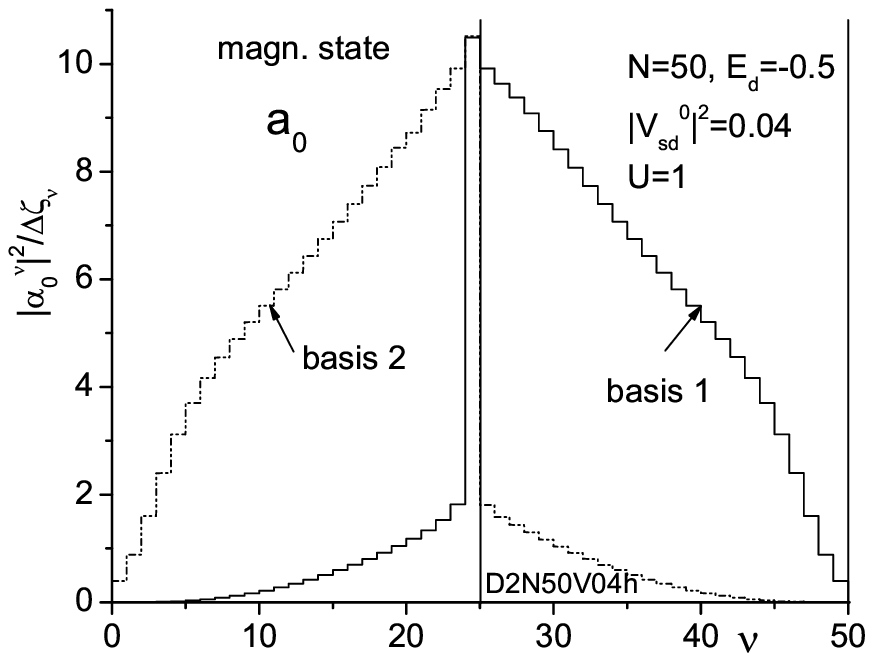}%
}%
\end{array}
\\
&
\begin{tabular}
[c]{l}%
Fig.6: The coefficients $\alpha_{0\pm}^{\nu}$ of the FAIR states $a_{0+}%
^{\dag}$ and $a_{0-}^{\dag}$ are plotted in terms \\
of the states $c_{\nu}^{\dag}$ for the magnetic state $\Psi_{MS}$. The full
and the dashed curves \\
represent $a_{0+}^{\dag}$ and $a_{0-}^{\dag}$ respectively.\\
Fig.6b: The density distribution of the $p_{\nu}=$ $\left\vert \alpha_{0\pm
}^{\nu}\right\vert ^{2}/\Delta_{\nu}$ as a function of as a \\
function of $\nu.$%
\end{tabular}
\end{align*}

In Fig.6b the quantity $p_{\nu}$=$\left\vert \alpha_{0}^{\nu}\right\vert
^{2}/\Delta_{\nu}$ is plotted as a function $\nu$, where $\Delta_{\nu}=\left(
\zeta_{\nu+1}-\zeta_{\nu}\right)  $ is the width of the energy cell
$\mathfrak{C}_{\nu}$ of the Wilson state $\psi_{\nu}\left(  x\right)  $. The
quantity $p_{\nu}$ is a function of the energy $p_{\nu}=p\left(  \zeta\right)
,$ and $p\left(  \zeta\right)  d\zeta$ represents the weight of the original
basis states $\varphi_{k}\left(  x\right)  $ in the energy window $\left(
\zeta,\zeta+d\zeta\right)  $ to the FAIR state $a_{0}^{\dag}$. This weight is
plotted in Fig.6b for the magnetic state as a function of $\nu$. It would be
more natural to plot $p\left(  \zeta\right)  $ as a function of the energy.
But for $\nu$ close to $N/2$ the width of the energy cells $\mathfrak{C}_{\nu
}$ is less than $10^{-6}$ and any dependence of $p\left(  \zeta\right)  $ on
the energy cannot be resolved on a linear scale. The probability $p_{\nu}$
increases close to the Fermi energy. Again the contributions of $a_{0+}^{\dag
}$ and $a_{0-}^{\dag}$ resemble mirror images.

In Fig.7a and Fig.7b the corresponding plots are shown for the magnetic
component of the singlet state. While the overall shape of the coefficients in
Fig.7a resembles that in Fig.6 the weight $p_{\nu\pm}=\left\vert \alpha_{0\pm
}^{\nu}\right\vert ^{2}/\Delta_{\nu}$ close to the Fermi surface is very
different for the singlet state and the magnetic state. While the two
probabilities are mirror images for the magnetic case and reach a maximum
value of about 10 one observes in the singlet state a maximum of about 400 and
the weights $p_{\nu\pm}$ in $a_{0+}^{\dag}$ and $a_{0-}^{\dag}$ are
essentially identical and not mirror images. The magnetic component of the
singlet state is in a subtle way different from the magnetic state.
\begin{align*}
&
\begin{array}
[c]{cc}%
{\includegraphics[
height=2.1602in,
width=2.7073in
]%
{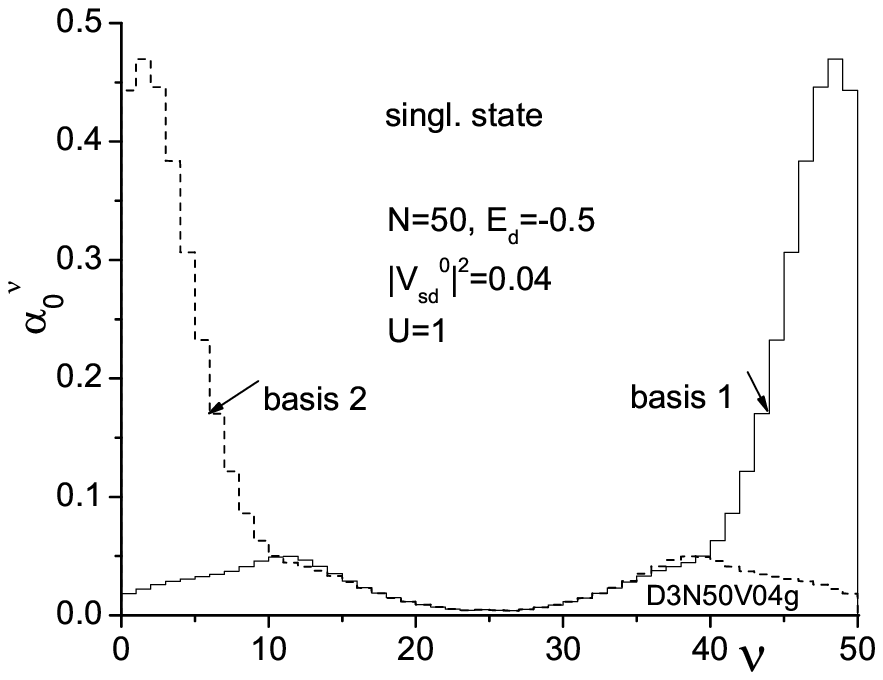}%
}%
&
{\includegraphics[
height=2.122in,
width=2.7978in
]%
{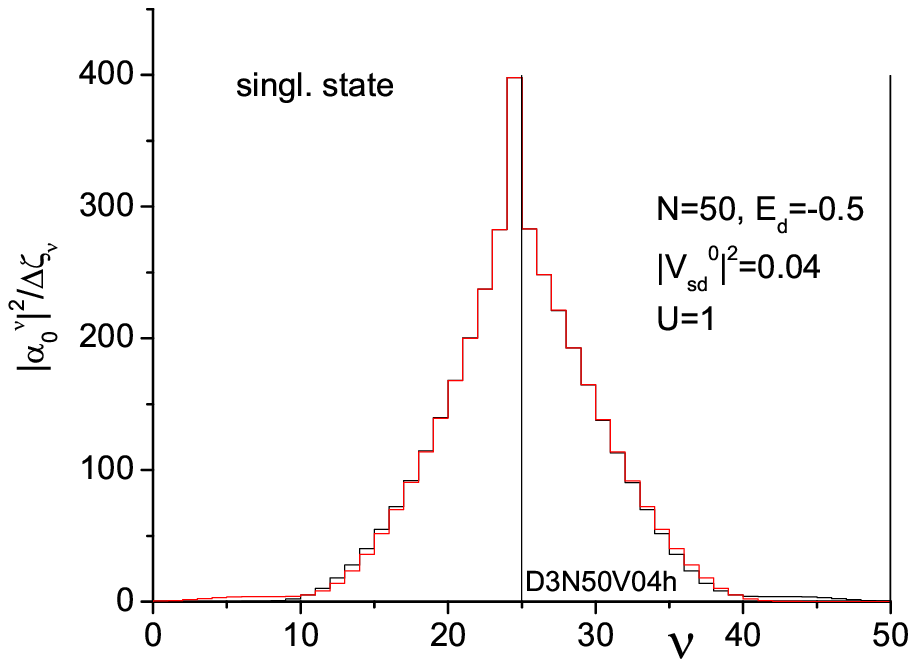}%
}%
\end{array}
\\
&
\begin{tabular}
[c]{l}%
Fig.7a: The coefficients $\alpha_{0\pm}^{\nu}$ of the FAIR states
$a_{0+}^{\dag}$ and $a_{0-}^{\dag}$ in terms of\\
the Wilson states $c_{\nu}^{\dag}$ of the magnetic component $\overline
{\Psi_{MS}}$ of the singlet state.\\
Fig.7b: The density distribution of the $p_{\nu}=$ $\left\vert \alpha_{0\pm
}^{\nu}\right\vert ^{2}/\Delta_{\nu}$ as a function of \ \\
the cell number $\nu.$%
\end{tabular}
\end{align*}

It is interesting to note that the spatial densities of the two FAIR states
$a_{0+}^{\dag}$ and $a_{0-}^{\dag}$ (after averaging over Friedel
oscillations) are identical with an accuracy of four digits. This applies for
the magnetic and the singlet state. However, their energy expectation values
are essentially opposite equal. The values for $\left\langle a_{0+}^{\dag}%
\Phi_{0}\left\vert H\right\vert a_{0+}^{\dag}\Phi_{0}\right\rangle $ are
rather similar in the magnetic state $\left(  \thickapprox+0.288\right)  $ and
the singlet ground state $\left(  +0.2865\right)  $ of the FA impurity.

\subsection{Kondo Impurity}

If one increases the Coulomb potential in the Friedel-Anderson impurity (and
lowers the d-state energy $E_{d}$) then the coefficients $A_{s,s}$ and
$A_{d,d}$ converge towards zero. This is the limit of the Kondo impurity. By
means of the Schrieffer-Woulff transformation one can express the
s-d-interaction as an exchange interaction of the form $2J\mathbf{s\cdot S}$.
In this limit the compact ground state $\Psi_{K}$ takes the form%
\begin{align}
\Psi_{K}  &  =\left[  A_{s,d}a_{0+\uparrow}^{\dag}d_{\downarrow}^{\dag
}+A_{d,s}d_{\uparrow}^{\dag}a_{0-\downarrow}^{\dag}\right]
{\textstyle\prod\limits_{i=1}^{n-1}}
a_{i+\uparrow}^{\dag}%
{\textstyle\prod\limits_{j=1}^{n-1}}
a_{j-\uparrow}^{\dag}\Phi_{0}\label{Psi_K}\\
&  +\left[  A_{d,s}a_{0-\uparrow}^{\dag}d_{\downarrow}^{\dag}+A_{s,d}%
d_{\uparrow}^{\dag}a_{0+\downarrow}^{\dag}\right]
{\textstyle\prod\limits_{i=1}^{n-1}}
a_{i-\uparrow}^{\dag}%
{\textstyle\prod\limits_{j=1}^{n-1}}
a_{j+\uparrow}^{\dag}\Phi_{0}\nonumber
\end{align}

Since equ.(\ref{Psi_K}) is a reduced form of equ. (\ref{Psi_SS}) the
calculation of the integrated net density is analogous to the previous
calculation. The Kondo singlet state is composed of two magnetic components
represented by the first and the second terms. Note that the effective moment
of these magnetic building blocks is less than $1\mu_{B}$.

We investigate the Kondo ground state for the exchange parameter $J=0.1.$ In
Fig.8 the net integrated density is plotted versus the logarithm of the
distance $r=2^{l}$ for the magnetic component with the d-spin up. One
recognizes that a spin polarized "cloud" extends up to a distance of
$r=2^{14.7}$ from the Kondo impurity.
\begin{align*}
&
{\includegraphics[
height=2.9141in,
width=3.6538in
]%
{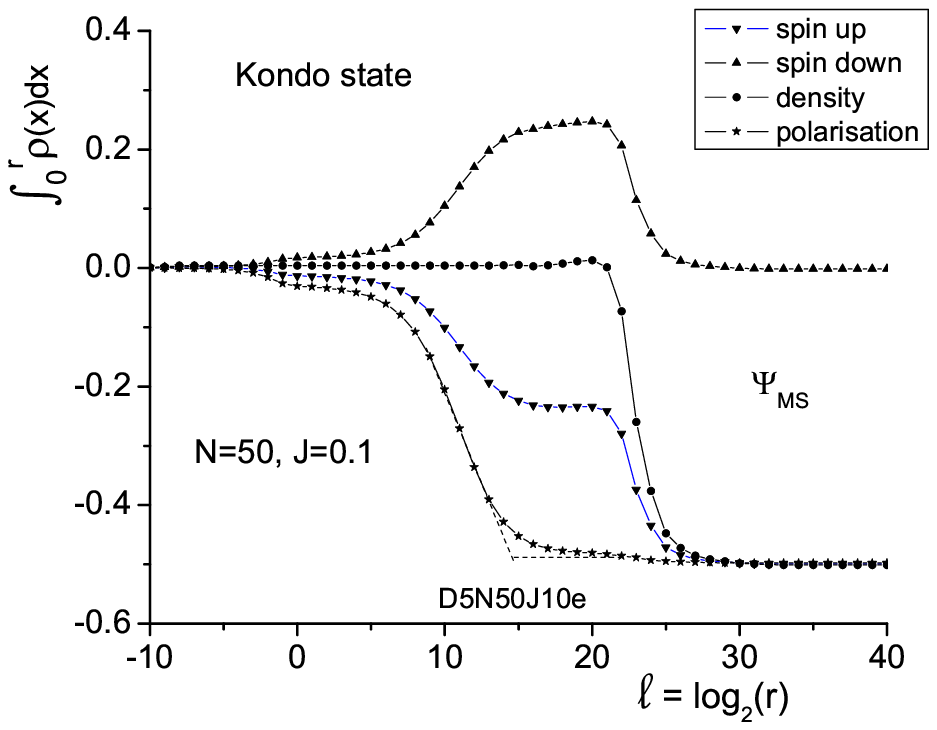}%
}%
\\
&
\begin{tabular}
[c]{l}%
Fig.8: The net integrated density $\int_{0}^{r}\rho\left(  x\right)  dx$
within a distance $r=2^{l}$\\
from the d-spin up component of the impurity$.$ Shown are the spin up, spin\\
down components as well as the total density and the polarization.
\end{tabular}
\end{align*}

Next we want to study the extent of the Kondo cloud as a function of the Kondo
energy. In a recent investigation we obtained for the coupling of $J=0.1$ a
relaxed singlet-triplet excitation energy of $\Delta E_{st}\left(  0.1\right)
=2.35\times10^{-5}$. By reducing the coupling to $J=0.08$ we observed an
$\Delta E_{st}\left(  0.08\right)  =1.37\times10^{-6}$. The ratio of the two
energies is equal to $17.1$.

Fig.9 shows the integrated net s-electron density for a Kondo impurity with
$\ J=0.01$. The two polarization curves (stars) in Fig.8 and Fig.9 are very
nicely parallel. Therefore the relative shift of the two curves can be
determined very accurately. It is equal to $4.1$ in terms of the $\log_{2}%
$-scale, which corresponds to a length ratio of $2^{4.1}=\allowbreak
17.\,\allowbreak1$. This is a very good agreement with the ratio of $17.1$ of
the two Kondo energies. Therefore we can confirm that the extent of the
screening cloud scales with the Kondo energy. Kondo energy and extension in
real space are reciprocal.%

\begin{align*}
&
{\includegraphics[
height=3.2511in,
width=4.1079in
]%
{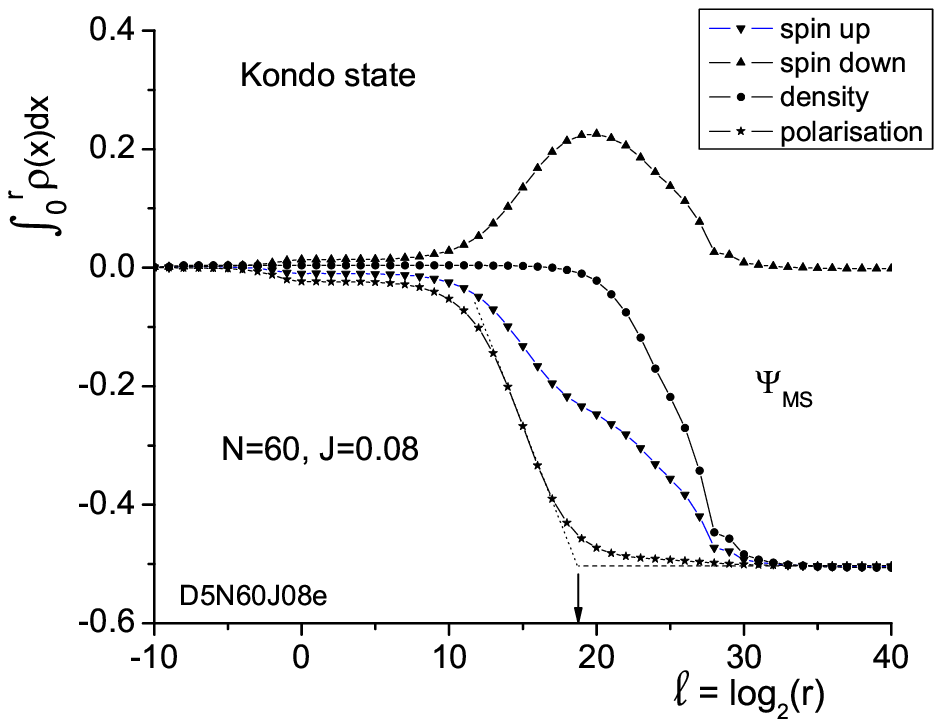}%
}%
\\
&
\begin{tabular}
[c]{l}%
Fig.9: The net integrated density $\int_{0}^{r}\rho\left(  x\right)  dx$
within a distance $r=2^{l}$\\
from the magnetic component of the impurity$.$ Shown are the spin up, spin\\
down components as well as the total density and the polarization. The\\
polarization extends over a distance of $r=2^{18.8}\thickapprox$
$4.\,\allowbreak6\times10^{5}$ $\left[  \lambda_{F}/2\right]  .$%
\end{tabular}
\end{align*}

If in Fig.9 we extrapolate the range of the Kondo cloud as the intersection
between the linear decrease and the saturation at the right then we obtain an
intersection of $l=18.8$ or a range of $2^{18.8}$ (in units of $\lambda
_{F}/2)$. In the reduced length units of $\lambda_{F}/2$ and the energy units
of $\varepsilon_{F}$ the Kondo length equ. (\ref{K_xi}) has the values
$\xi_{K}=2/\left(  \pi\varepsilon_{K}\right)  $ where $\varepsilon_{K}$ is the
Kondo energy in units of $\varepsilon_{F}$. For the parameters of Fig.9
($J=0.08)$ the Kondo energy has the value $\varepsilon_{K}=\Delta
E_{st}=1.37\times10^{-6}$. This yields $\xi_{K}=4.\,\allowbreak6\times
10^{5}\thickapprox2^{18.8}$. The perfect agreement with the numerical
extrapolation is probably lucky because the equation (\ref{K_xi}) for the
Kondo length is intended as an order of magnitude.

\section{Conclusion}

A recently developed compact solution for the singlet state of the
Friedel-Anderson and the Kondo impurity is used to investigate the old
question of a Kondo cloud in the Kondo ground state. Wilson's states with an
exponentially decreasing frame of energy cells towards the Fermi level and a
linear dispersion relation are expressed in free electron waves integrated
over the width of their energy cells. A rotational iteration process (which
has been described in previous papers) is used to obtain the optimal ground
state. The singlet state of the Friedel-Anderson impurity consists of eight
Slater states. They are divided into two groups of four, where each group
represents one magnetic component of the singlet state. Two criteria yield the
same splitting of the eight Slater states: (i) Each group transforms into the
other one by inverting all s- and d-spins, (ii) The two groups combine
symmetrically in the singlet and anti-symmetrically in the triplet state.

After identifying the magnetic components, the real space wave function is
evaluated. Since the spatial range of the Wilson states varies in the range
between a Fermi wave length $\lambda_{F}$ and $2^{N/2}\lambda_{F}$ ($N$
$\thickapprox50$ is the number of Wilson states used in the evaluation) one
has to deal with electron densities which vary be a factor of $2^{25}$
$\thickapprox3\times10^{7}$. Therefore it is favorable to integrate over the
net s-electron density from the impurity outwards. The resulting curves yield
the spatial range over which the spin densities of the s-electrons contribute
to the polarization about the impurity.

If one wants to obtain the polarization about the impurity then one has to
calculate the difference of spin up and down s-electron densities with and
without the impurity. The simple and exact solution of the Friedel resonance
for spinless electrons is used to test the method and interpret the results.
It turns out that for each Slater state the behavior of the (integrated)
density and polarization can be divided into two regions. For $r>2^{N/2}$ the
system follows the sum rule which is imposed by the number of s- and d-states
in the multi-electron Slater state. For $r<2^{N/2}$ one obtains the physical
densities and polarizations. The behavior in this range does not change when
$N$ is increased. If the d-state in the Friedel resonance is set at the Fermi
level then one observes in real space that the range of the s-electron cloud
is proportional to $1/\left|  V_{sd}^{0}\right|  ^{2}$.

For the magnetic state $\Psi_{MS}$ of the Friedel-Anderson impurity one
observes essentially no magnetic s-polarization in the vicinity of the d-impurity.

For the magnetic component $\overline{\Psi_{MS}}$ of the singlet state in the
Friedel-Anderson impurity one observes an s-polarization cloud which screens
the spin (magnetic moment) of the d-electron. The range of this polarization
cloud is investigated in detail for the Kondo impurity. The range is inversely
proportional to the Kondo energy. The latter was obtained in a previous
investigation as the energy difference between the singlet state and the
relaxed triplet state energy. The absolute value of the range of the
polarization cloud agrees surprisingly well with the Kondo length.

The different screening behavior in the magnetic state $\Psi_{MS}$ and the
singlet state $\Psi_{SS}$ is due to subtle differences in the composition of
the FAIR states $a_{0+}^{\dag}$ and $a_{0-}^{\dag}$. In the singlet state the
FAIR states have a much larger weight (of the original basis states
$\varphi_{k}$) very close to the Fermi energy. It is remarkable that the two
FAIR states $a_{0+}^{\dag}$ and $a_{0-}^{\dag}$, which are far apart in their
energy, have essentially an identical density in real space (after averaging
over the Friedel oscillations).

In the present paper a very simple energy band and dispersion relation is used
in analogy to Wilson's work. This simplifies the numerical evaluation
dramatically. However, the spatial dependence can be evaluated for an
arbitrary s-band with an energy-dependent density of states and
s-d-interaction. The author generalized the definition of the Wilson states
\cite{B159} for this case.\ The numerical work would, however, be much more
extensive. (The spin of the impurity is still restricted to $1/2$).

It should be emphasized that the extent of the Kondo cloud calculated here
applies only when the mean free path of the conduction electrons is larger
than the Kondo length. Also the surface of the host should be at least this
distance away from the impurity. This does, however, not mean that for smaller
sample size the Kondo effect is suppressed. As long as there are sufficiently
many electron states within the Kondo resonance at the Fermi level which
couple to the impurity one obtains the full Kondo effect \cite{B59}.

Finally it should be noted that the extent of the electron density in real
space is a detector for a resonance in energy. The spatial extension $\xi$ and
the resonance width $\Delta$ are reciprocal and given by the relation
$\xi\Delta\thickapprox\hbar v_{F}$. The Wilson states act here as a magnifying
glass close to the Fermi energy because their energy width $\Delta_{\nu}$
decreases as $2^{-\nu}$ towards the Fermi level. This is particularly
demonstrated for the Friedel impurity in Fig.2. The extent of the net
integrated density is only observed when the energy of the d-state lies at the
Fermi level. Otherwise it cannot be detected. For example with $E_{d}=-0.5$
the the resolution of the Wilson states is only between 0.5 and 0.25 which is
not sufficient to detect the Friedel resonance. One has to split the Wilson
states close the resonance to make it visible. Therefore the Wilson states can
be used as a quasi-experimental spectroscope. 

\appendix{}

\section{The FAIR Method}

The Hamiltonian of the Friedel-Anderson impurity is given in equ.
(\ref{hfa0}). One obtains the mean-field Hamiltonian from equ.(\ref{hfa0}) by
replacing $n_{d+}n_{d-}$ =%
$>$%
$n_{d+}\left\langle n_{d-}\right\rangle $ $+\left\langle n_{d+}\right\rangle
n_{d-}$ $-\left\langle n_{d+}\right\rangle \left\langle n_{d-}\right\rangle $.
After adjusting $\left\langle n_{d+}\right\rangle $ and $\left\langle
n_{d-}\right\rangle $ self-consistently one obtains two Friedel resonance
Hamiltonians with a spin-dependent energy of the $d_{\sigma}$-state:
$E_{d,\sigma}$ $=E_{d}+U\left\langle n_{d,-\sigma}\right\rangle $.
\[
H_{mf}=%
{\textstyle\sum_{\sigma}}
\left\{  \sum_{\nu=1}^{N}\varepsilon_{\nu}c_{\nu\sigma}^{\dag}c_{\nu\sigma
}+E_{d\sigma}d_{\sigma}^{\dag}d_{\sigma}+\sum_{\nu=1}^{N}V_{sd}(\nu
)[d_{\sigma}^{\dag}c_{\nu\sigma}+c_{\nu\sigma}^{\dag}d_{\sigma}]\right\}
\]
The mean-field wave function is a product of two Friedel ground states for
spin up and down $\Psi_{mf}=\Psi_{F\uparrow}\Psi_{F\downarrow}$ .

It has been shown \cite{B91}, \cite{B92} that the \textbf{exact} $n$-particle
ground state of the Friedel Hamiltonian can be expressed as by the sum of two
Slater states, in which either the $d^{\dag}$-state or a state $a_{0}^{\dag}$
is multiplied with the same $\left(  n-1\right)  $-s-electron state $%
{\textstyle\prod\limits_{i=1}^{n-1}}
a_{i}^{\dag}\Phi_{0}$ (see equ. (\ref{Psi_F})). The state $a_{0}^{\dag}=%
{\textstyle\sum_{\nu}}
\alpha_{0,\nu}c_{\nu}^{\dag}$ is a localized state whose components
$\alpha_{0,\nu}$ can be obtained analytically \cite{B92} or by variation
\cite{B91}.%

\begin{align*}
&
{\includegraphics[
height=2.406in,
width=4.8003in
]%
{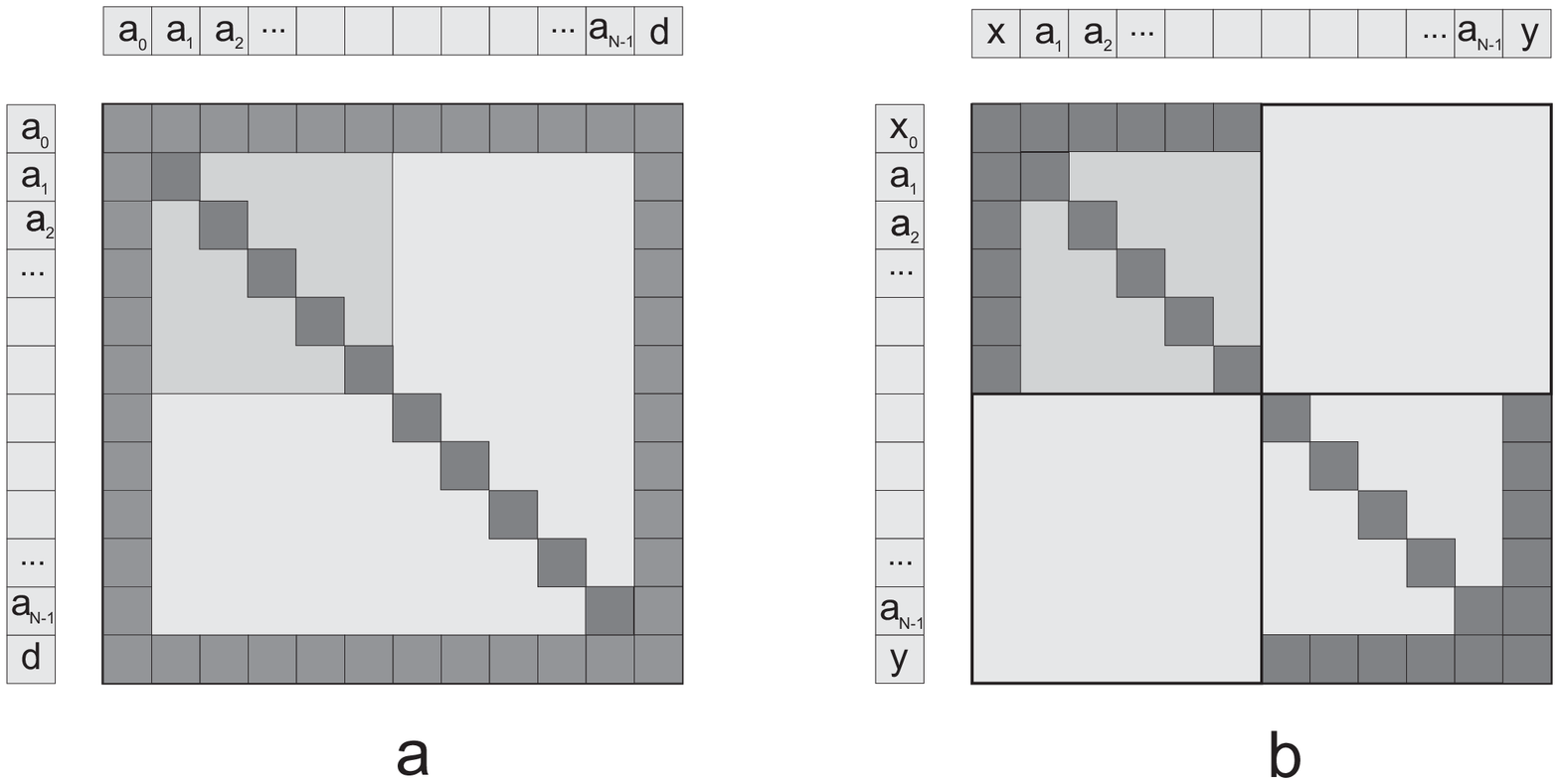}%
}%
\\
&
\begin{tabular}
[c]{l}%
Fig.10a: The matrix of the Friedel Hamiltonian in the basis $\left\{
a_{i}^{\dag}\right\}  ,d^{\dag}$.\\
Fig.10b: The two basis vectors $a_{0}^{\dag},d^{\dag}$ are rotated into
$x^{\dag}=\left(  Aa_{0}^{\dag}+Bd^{\dag}\right)  $\\
and $y^{\dag}=\left(  Ba_{0}^{\dag}-Ad^{\dag}\right)  $. For the optimal
$a_{0}^{\dag}$ the matrix is block-diagonal.
\end{tabular}
\end{align*}
In Fig.10a the matrix of the Friedel Hamiltonian in the basis $\left\{
a_{i}^{\dag}\right\}  ,d^{\dag}$ is shown. Only the diagonal elements and the
elements along the edges are non-zero. The inner $\left(  N-1\right)
\times\left(  N-1\right)  $-sub-matrix is diagonalized. This way the state
$a_{0}^{\dag}$ determines uniquely the full basis $\left\{  a_{i}^{\dag
}\right\}  $. One recognizes that the state $a_{0}^{\dag}$ represents an
artificially inserted Friedel resonance state, with similar properties as the
state $d^{\dag}$. Therefore I call $a_{0}^{\dag}$ a "Friedel artificially
inserted resonance" state or \textbf{FAIR}-state. The use of the FAIR-states
is at the heart of my approach to the FA- and Kondo impurity problem.
Therefore I call this approach the \textbf{FAIR }method.

The structure of the Hamiltonian in Fig.10a can be obtained for an arbitrary
state $a_{0}^{\dag}$. There is, however, one optimal state $a_{0}^{\dag}$
which yields the exact n-electron eigenstate $\Psi_{F}$ and for which the
energy expectation value $\left\langle \Psi_{F}\left\vert H_{F}\right\vert
\Psi_{F}\right\rangle $ has a minimum. With this state the two basis vectors
$a_{0}^{\dag},d^{\dag}$ can be rotated within their plane into the orthogonal
states $x^{\dag}=\left(  Aa_{0}^{\dag}+Bd^{\dag}\right)  $ and $y^{\dag
}=\left(  Ba_{0}^{\dag}-Ad^{\dag}\right)  $. For the optimized $a_{0}^{\dag}$
the resulting Hamiltonian is block-diagonal. This is shown in Fig.10b. The
upper left block contains the occupied states. Since there is no
matrix-element between the occupied and the empty block the product state
$x^{\dag}a_{1}^{\dag}a_{2}^{\dag}..a_{n-1}^{\dag}\Phi_{0}$ $=\left(
Aa_{0}^{\dag}+Bd^{\dag}\right)
{\textstyle\prod\limits_{i=1}^{n-1}}
a_{i}^{\dag}\Phi_{0}$ is an eigenstate.

For the two Friedel states in the mean-field wave function I use the form of
equ. (\ref{Psi_F}) and obtain for the mean-field solution
\begin{equation}
\Psi_{mf}=\left[  \left(  A_{+}a_{0+\uparrow}^{\dag}+B_{+}d_{\uparrow}^{\dag
}\right)  \prod_{i=1}^{n-1}a_{i+\uparrow}^{\dag}\right]  \left[  \left(
A_{-}a_{0-\downarrow}^{\dag}+B_{-}d_{\downarrow}^{\dag}\right)  \prod
_{i=1}^{n-1}a_{i-\downarrow}^{\dag}\right]  \Phi_{0} \label{Psi_mf}%
\end{equation}
where $\left\{  a_{i+}^{\dag}\right\}  $ and $\left\{  a_{i-}^{\dag}\right\}
$ are two (different) bases of the $N$-dimensional Hilbert space. This
solution can be rewritten as equation (\ref{Psi_MS}).

In the mean-field solution $\Psi_{mf}$ the coefficients $A_{\alpha,\beta}$ are
restricted by two conditions $A_{\pm}^{2}+B_{\pm}^{2}=1$ (here $A_{s,s}%
=A_{+}A_{-}$ etc). Therefore this state does not describe well the correlation effects.

In contrast the state (\ref{Psi_MS}) opens a wide playing field for improving
the solution: (i) The FAIR states $a_{0+}^{\dag}$ and $a_{0-}^{\dag}$ can be
individually optimized, each one defining a whole basis $\left\{  a_{i\pm
}^{\dag}\right\}  $ (which yields a Hamiltonian of the form in Fig.10a) and
(ii) the coefficients $A_{\alpha,\beta}$ can be optimized as well fulfilling
only the normalization condition $A_{s,s}^{2}+A_{d,s}^{2}+A_{s,d}^{2}%
+A_{d,d}^{2}=1$. This yields a much better treatment of the correlation
effects. The resulting state is denoted as the (potentially) magnetic state
$\Psi_{MS}$. The magnetic state $\Psi_{MS}$ has the same form as the mean
field solution $\Psi_{mf}$; the only difference is that its components are
optimized for the Friedel-Anderson Hamiltonian. The optimization procedure is
described in detail in ref. \cite{B152}.

The magnetic state $\Psi_{MS}$ is used as the building block for the singlet
state. It form is given in equ. (\ref{Psi_SS}).

\section{Wilson's states}

Wilson considered an s-band with constant density of states and the Fermi
energy in the center of the band. By measuring the energy from the Fermi level
and dividing all energies by the Fermi energy Wilson obtained a band ranging
from $-1$ to $+1.$ To treat the electrons close to the Fermi level at
$\zeta=0$ as accurately as possible he divided the energy interval $\left(
-1:0\right)  $ at energies of $-1/2,-1/4,-1/8,..$ i.e. $\zeta_{\nu}=-1/2^{\nu
}.$ This yield energy cells $\mathfrak{C}_{\nu}$ ranging from $-1/2^{\nu}$
to$\ -1/2^{\nu+1}$ with the width $\Delta_{\nu}$ $=\zeta_{\nu+1}-\zeta_{\nu}$
$=1/2^{\nu+1}$.

Wilson rearranged the quasi-continuous original electron states $\varphi
_{k}\left(  x\right)  $ in such a way that only one state within each cell
$\mathfrak{C}_{\nu}\ $had a finite interaction with the impurity. Assuming
that the interaction of the original electron states $\varphi_{k}\left(
x\right)  $ with the impurity is $k$-independent this interacting state in
$\mathfrak{C}_{\nu}$ had the form%
\[
\psi_{\nu}\left(  x\right)  =%
{\textstyle\sum_{\mathfrak{C}_{\nu}}}
\varphi_{k}\left(  x\right)  /\sqrt{Z_{\nu}}%
\]
where $Z_{\nu}$ is the total number of states $\varphi_{k}\left(  x\right)  $
in the cell $\mathfrak{C}_{\nu}$ ($Z_{\nu}=Z\left(  \zeta_{\nu+1}-\zeta_{\nu
}\right)  /2,$ $Z$ is the total number of states in the band). There are
$\left(  Z_{\nu}-1\right)  $ additional linear combinations of the states
$\varphi_{k}$ in the cell $\mathfrak{C}_{\nu}$ but they have zero interaction
with the impurity and were ignored by Wilson as they will be within this paper.

The interaction strength of the original basis states $\varphi_{k}\left(
x\right)  $ with the d-impurity is assumed to be a constant, $v_{sd}$. Then
the interaction between the d-state and the Wilson states $\psi_{\nu}\left(
x\right)  $ is given by $V_{sd}\left(  \nu\right)  =V_{sd}^{0}\sqrt{\left(
\zeta_{\nu+1}-\zeta_{\nu}\right)  /2}$ where $\left\vert V_{sd}^{0}\right\vert
^{2}=$ $%
{\textstyle\sum_{k}}
\left\vert v_{sd}\right\vert ^{2}=$ $%
{\textstyle\sum_{\nu}}
\left\vert V_{sd}\left(  \nu\right)  \right\vert ^{2}$. 

\subsection{The wave function of Wilson's states in real space}

For the discussion of the wave functions in real space one has to look a bit
closer. We assume a linear dispersion relation between energy and momentum, a
constant density of states and a constant amplitude at $x=0.$ Then the
interaction between the original basis states $\varphi_{k}\left(  x\right)  $
and the d-impurity will be constant for all k-states. These assumption are the
same as in Wilson's treatment of the Kondo impurity. We define the wave
functions $\varphi_{k}\left(  x\right)  $ is such a way that the results apply
for the impurity problem in one, two and three dimensions.

\subsubsection{One-dimensional case}

Let us start with the one dimensional problem. Here we have the impurity at
the position zero and the conduction electrons are located in the range
between $0$ and $L.$ The wave functions $\varphi_{k}\left(  x\right)  $ have
the form $\varphi_{k}\left(  x\right)  =\sqrt{2/L}\cos\left(  kx\right)  $.
There is another set of eigenstates $\overline{\varphi_{k}\left(  x\right)
}=\sqrt{2/L}\sin\left(  kx\right)  $. These states don't interact with the
impurity at the origin. Therefore they don't have any bearing on the impurity problem.

\subsubsection{Three-dimensional case}

In three dimensions the free electron states can be expressed as $\varphi
_{k}\left(  x\right)  =Y_{l}^{m}\left(  \theta,\phi\right)  j_{l}\left(
kx\right)  $ where $Y_{l}^{m}$ is a spherical harmonics and $l,m$ are the
angular momentum and magnetic quantum numbers. $j_{l}\left(  kx\right)  $ is a
spherical Bessel function. Its long range behavior is given by $\left(
1/r\sqrt{2\pi L}\right)  \sin\left(  kx-l\pi/2\right)  $. Only the states with
the same $l$ as the impurity couple to the impurity. All the other states for
different $l$ belong to the group of inert states $\overline{\varphi
_{k}\left(  x\right)  }$.

If one calculates the density of the wave function, integrates in the
three-dimensional case over the spherical surface $4\pi r^{2}$ and averages
over short range (Friedel) oscillations then one obtains in the

\begin{itemize}
\item one-dimensional case: $\left(  2/L\right)  \overline{\cos^{2}\left(
kx\right)  }=2/L$

\item three-dimensional case: $\rho_{k}\left(  x\right)  =\left(  2/L\right)
\overline{\sin^{2}\left(  kx-l\pi/2\right)  }=2/L$
\end{itemize}

In both cases one obtains essentially the same density. Therefore it is
sufficient to use the one-dimensional approach for calculating the density of
a Kondo cloud. It is equivalent to the 3-dimensional case integrated over the
spherical surface.

\subsubsection{The wave functions in one dimension}

While the energy is measured in units of the Fermi energy the momentum will be
measured in units of the Fermi wave number. We assume a linear dispersion
relation for $0\leq k\leq2$ with
\[
\zeta=\left(  k-1\right)
\]
The (almost) continuous states $\varphi_{k}$ are given as
\[
\varphi_{k}\left(  x\right)  =\sqrt{\frac{2}{L}}\cos\left(  \pi kx\right)
\]
where $L$ is the length of the one-dimensional box. Since $k$ is measured in
units of $k_{F}$ the coordinate $x$ gives the position in units of
$\lambda_{F}/2$ where $\lambda_{F}$ is the Fermi wave number. The boundary
condition $\cos\left(  \pi kL\right)  =0$ yields $k=\left(  \lambda
+1/2\right)  /L$. (The maximal value of $\lambda$ is $2L,$ since $k$ is
dimensionless then $L$ is also dimensionless). Therefore we have $Z=2L$ states
in the full band of width $2$.

To obtain the Wilson state we have to sum the states $\varphi_{k}\left(  \pi
kx\right)  $ over all states within an energy cell. If the cell ranges from
$\left(  \zeta_{\nu}:\zeta_{\nu+1}\right)  $ corresponding to a $k$-range
$\left(  1+\zeta_{\nu}\right)  <k<\left(  1+\zeta_{\nu+1}\right)  $ then we
represent all the states in this energy interval by%
\[
\psi_{\nu}\left(  x\right)  =\frac{1}{\sqrt{\left(  \zeta_{\nu+1}-\zeta_{\nu
}\right)  L}}%
{\textstyle\sum_{1+\zeta_{\nu}<k<1+\zeta_{\nu+1}}}
\sqrt{\frac{2}{L}}\cos\left(  \pi kx\right)
\]
From $Z_{\nu}=L\left(  \zeta_{\nu+1}-\zeta_{\nu}\right)  $ states we have
(according to Wilson) constructed one state $\psi_{\nu}\left(  x\right)  $
which couples to the impurity. The same procedure yields $\left(  Z_{\nu
}-1\right)  $ additional linear combinations of $\varphi_{k}\left(  x\right)
$ which do not couple with the impurity at the origin. We denote these states
as $\overline{\overline{\varphi_{k}\left(  x\right)  }}$. They are as inert to
the impurity as the states $\overline{\varphi_{k}\left(  x\right)  }$ and will
be included in the quasi-vacuum.

The wave function of the state of the states $c_{\nu}^{\dag}$ has the form for
$\nu<N/2$%
\[
\psi_{\nu}\left(  x\right)  =\frac{2\sqrt{2}}{\sqrt{\left(  \zeta_{\nu
+1}-\zeta_{\nu}\right)  }}\frac{\sin\left(  \frac{\pi x\left(  \zeta_{\nu
+1}-\zeta_{\nu}\right)  }{2}\right)  }{\pi x}\cos\left(  \frac{\pi x\left(
2+\zeta_{\nu}+\zeta_{\nu+1}\right)  }{2}\right)
\]
For the exponential energy scale this yields for $\nu<N/2-1$%
\begin{equation}
\psi_{\nu}\left(  x\right)  =2\sqrt{2^{\nu+2}}\frac{\sin\left(  \pi x\frac
{1}{2^{\nu+2}}\right)  }{\pi x}\cos\left(  \pi x\left(  1-\frac{3}{2^{\nu+2}%
}\right)  \right)  \label{psi_nu}%
\end{equation}
Similarly one obtains for in the positive energy range
\[
\psi_{N-1-\nu}\left(  x\right)  =2\sqrt{2^{\nu+2}}\frac{\sin\left(  \frac
{1}{2^{\nu+2}}\pi x\right)  }{\pi x}\cos\left(  \pi x\left(  1+\frac{3}%
{2^{\nu+2}}\right)  \right)
\]

The two wave functions $\psi_{N/2-1}$ and $\psi_{N/2}$ are special because
their $k$-range is the same as their neighbors $\psi_{N/2-2}$ and
$\psi_{N/2+1}$ All four states close to the Fermi level have the same
$k$-range of $2^{-N/2-1}.$ One has to pay special attention to this complication.

\subsection{Density of the Wilson states in real space}

The density of the wave function $\psi_{\nu}\left(  x\right)  $ is given by%
\[
\left\vert \psi_{\nu}\left(  x\right)  \right\vert ^{2}=\frac{8}{\left(
\zeta_{\nu+1}-\zeta_{\nu}\right)  }\frac{\sin^{2}\left(  \frac{\pi x\left(
\zeta_{\nu+1}-\zeta_{\nu}\right)  }{2}\right)  }{\left(  \pi x\right)  ^{2}%
}\cos^{2}\left(  \frac{\pi x\left(  2+\zeta_{\nu}+\zeta_{\nu+1}\right)  }%
{2}\right)
\]

The density of a single state $\psi_{\nu}\left(  x\right)  $ is given by the
square of the function $\psi_{\nu}\left(  x\right)  $ in equ. (\ref{psi_nu}).
This density has a fast oscillating contribution which yields the Friedel
oscillations. We are here interested in the density on a much larger scale and
average over the fast oscillation (which has a period of 1 in units of
$\lambda_{F}/2)$. Then we obtain for the Wilson states
\begin{equation}
\rho_{\nu}^{0}\left(  x\right)  =\left\vert \psi_{\nu}\left(  x\right)
\right\vert ^{2}=2^{\nu+3}\frac{\sin^{2}\left(  \pi x\frac{1}{2^{\nu+2}%
}\right)  }{\pi x}dx \label{ro_nu}%
\end{equation}
In the numerical calculation we will use (most of the time) $N=50$ Wilson
states. From equ. (\ref{ro_nu}) one recognizes that the density $\rho_{\nu
}^{0}$ of a state $\psi_{\nu}$ lies roughly in the interval between $2^{\nu
-2}$ and $2^{\nu+2}$ (in units of $\lambda_{F}/2$). Since for negative
energies $\nu$ takes the values from $0\ $to $\left(  N/2-1\right)  $ the
different wave functions $\psi_{\nu}\left(  x\right)  $ have very different
spatial ranges and therefore very different densities, the lowest being of the
order of $2^{-25}<3\times10^{-8}$. This means that it is not useful to
calculate the density as a function of $x$ because this density varies over a
range of $2^{25}$. Instead, we use the integrated density, integrated from $0$
to $r$.%
\begin{align*}
q_{\nu}^{0}\left(  r\right)   &  =\int_{0}^{r}\left\vert \psi_{\nu}\left(
x\right)  \right\vert ^{2}dx=2^{\nu+3}\int_{0}^{r}\frac{\sin^{2}\left(  \pi
x\frac{1}{2^{\nu+2}}\right)  }{\pi x}dx\\
&  =2\int_{0}^{\frac{r}{2^{\nu+2}}}\frac{\sin^{2}\left(  \pi u\right)
}{\left(  \pi u\right)  ^{2}}du
\end{align*}
One realizes that a single integral yields the integrated density for (almost)
all wave function $\psi_{\nu}\left(  x\right)  $. The state $\psi_{0}\left(
\pi x\right)  $ lies roughly in the range between $2^{-2}<r<2^{2},$ i.e. the
integrated density $q_{\nu}\left(  r\right)  =\int_{0}^{r}\left\vert \psi
_{0}\left(  \pi x\right)  \right\vert ^{2}dx$ increases in this range from
10\% to 90\%. Therefore the states $\psi_{\nu}\left(  x\right)  $ and
$\psi_{N-1-\nu}\left(  \pi x\right)  $ lie in the range between $2^{\nu
-2}<r<2^{\nu+2}$. For a total of $N=50$ Wilson states the maximum range of the
wave functions is roughly $2^{N/2}=2^{25}$.

We may define as a ruler a linear array $I\left(  s\right)  $ where $s$ is an
integer$,$ $\left(  -N/2\leq s<N\right)  $ as%
\[
I\left(  s\right)  =2\int_{0}^{2^{s}}\frac{\sin^{2}\left(  \pi u\right)
}{\left(  \pi u\right)  ^{2}}du
\]
Then the integrated density of the state $\psi_{\nu}$ in the range from $0$ to
$2^{l}$ is given by
\[
q_{\nu}^{0}\left(  2^{l}\right)  =\int_{0}^{2^{l}}\left\vert \psi_{\nu}\left(
x\right)  \right\vert ^{2}dx=2\int_{0}^{\frac{2^{l}}{2^{\nu+2}}}\frac{\sin
^{2}\left(  \pi u\right)  }{\left(  \pi u\right)  ^{2}}du=I\left(
l-\nu-2\right)
\]
Then $q_{\nu}^{0}\left(  2^{l}\right)  $ gives the integrated density on an
exponential scale.

\subsubsection{Interference terms in the density}

The Wilson states $\psi_{\nu}\left(  x\right)  $ or $c_{\nu}^{\dag}$ represent
the free electron states in the impurity problem. After the interaction we
express the ground state in terms of new states $a_{i}^{\dag}=%
{\textstyle\sum_{\nu=0}^{N-1}}
\alpha_{i}^{\nu}c_{\nu}^{\dag}$. Its integrated density is given by%
\[
\overline{\rho}_{i}\left(  2^{l}\right)  =\int_{0}^{2^{l}}\left\vert
{\textstyle\sum_{\nu=0}^{N-1}}
\alpha_{i}^{\nu}\psi_{\nu}\left(  x\right)  \right\vert ^{2}dx
\]

The quadratic terms can be evaluated with the same ruler $I\left(  s\right)  $
as before. But this time one has in addition interference terms $\psi_{\nu
}\left(  x\right)  \psi_{\nu+\lambda}\left(  x\right)  $. These terms depend
on two parameters, $\nu$ and $\lambda$. So one needs for each $\lambda$ a
different ruler. Furthermore the interference terms depend on the sub-bands of
$\psi_{\nu}\left(  x\right)  $ and $\psi_{\nu+\lambda}\left(  x\right)  $. If
both lie either in the negative energy sub-band ($\nu,\nu+\lambda<N/2$) or in
the positive sub-band ($\nu,\nu-\lambda\geq N/2$) then one obtains one set of
rulers $I_{0}\left(  \nu,\lambda\right)  $ and if they lie in opposite
sub-bands then one obtains another set of rulers $I_{1}\left(  \nu
,\lambda\right)  $. As an example one obtains%
\[
I_{0}\left(  s,\lambda\right)  =2\sqrt{2^{\lambda}}\int_{0}^{2^{s}}\frac
{\sin\left(  \pi u\right)  \sin\left(  \pi\frac{u}{2^{\lambda}}\right)
}{\left(  \pi u\right)  ^{2}}\cos\left(  3\pi u\left(  1-\frac{1}{2^{\lambda}%
}\right)  \right)  du
\]
For $I_{1}\left(  \nu,\lambda\right)  $ one has to replace the minus sign in
the cosine function by a plus sign. Furthermore one has to treat the terms
where $\nu+\lambda=N/2-1$ separately because one state lies at the Fermi level
and has a different cell width.

\subsubsection{The net integrated density}

If we occupy all Wilson states below the Fermi level then we obtain $%
{\textstyle\prod\limits_{\nu=0}^{n-1}}
c_{\nu}^{\dag}\Phi_{0}$ with $n=N/2$ and $\Phi_{0}$ the vacuum state. This
state is not really the free electron ground state. To obtain the latter we
have to occupy the states $\overline{\varphi_{k}\left(  x\right)  }$ and
$\overline{\overline{\varphi_{k}\left(  x\right)  }}$. They don't interact
with the impurity but they are occupied. Therefore we define as quasi-vacuum
$\Phi_{0}^{\prime}$ the state in which all states $\overline{\varphi
_{k}\left(  x\right)  }$ and $\overline{\overline{\varphi_{k}\left(  x\right)
}}$ with $k<k_{F}=1$ are occupied. Then the ground state is $\Psi_{0}=$ $%
{\textstyle\prod\limits_{\nu=0}^{n-1}}
c_{\nu}^{\dag}\Phi_{0}^{\prime}$. This state has a constant electron density
in real space.

In the presence of the impurity the new ground state $\Psi_{new}=$ $%
{\textstyle\prod\limits_{i=0}^{n-1}}
a_{i}^{\dag}\Phi_{0}^{\prime}$ must also contain this quasi vacuum, i.e., the
non-interacting states must be occupied up to the Fermi level. Since the inert
states are occupied in $\Psi_{0}$ and $\Psi_{new}$ they cancel out when one
calculates the change in the electron density. The net density of the new
state $\Psi_{new}$ is the difference between $\rho\left(  \Psi_{new}\right)  $
and $\rho\left(  \Psi_{0}\right)  $. Since the inert states cancel out one can
ignore their existence during this calculation.

\end{document}